\documentclass{article}

\usepackage{hyperref}

\usepackage{arxiv}

\usepackage[utf8]{inputenc} 
\usepackage[T1]{fontenc}    
\usepackage{url}            
\usepackage{booktabs}       
\usepackage{amsfonts}       
\usepackage{nicefrac}       
\usepackage{microtype}      
\usepackage{doi}

\usepackage{graphicx}
\usepackage{subcaption}
\usepackage{amsmath} 
\usepackage{amssymb} 
\usepackage{breqn}
\usepackage{algorithm}
\usepackage{algorithmic}
\usepackage{float}
\usepackage{amsthm}
\usepackage{enumitem,xcolor}
\usepackage{cite}

\theoremstyle{definition}
\newtheorem{exmp}{Example}[]

\newcommand{\mbs}[1]{\boldsymbol{#1}}
\newcommand{\mbf}[1]{\mathbf{#1}}
\newcommand{\T}{\top}

\newcommand{\nlos}{\textsc{Nlos}}
\newcommand{\los}{\textsc{Los}}
\newcommand{\rmse}{\textsc{Rmse}}
\newcommand{\toa}{\textsc{Toa}}
\newcommand{\tdoa}{\textsc{Tdoa}}

\newcommand{\async}{\textsc{Tdst}}

\newcommand{\cdf}{\textsc{cdf}}

\newcommand{\corr}{\epsilon}
\newcommand{\corrub}{\widetilde{\corr}}
\newcommand{\N}{N}
\newcommand{\Na}{N_{a}}
\newcommand{\parm}{\mbs{\theta}}
\newcommand{\loc}{\mbs{x}}
\newcommand{\rcv}{0}

\newcommand{\obsc}{z}
\newcommand{\obsvec}{\mbs{\obsc}}
\newcommand{\Nobs}{n}

\newcommand{\delay}{\delta}

\newcommand{\M}{\mbf{M}}

\newcommand{\rhovec}{\mbs{\rho}}
\newcommand{\R}{\mathbb{R}}
\newcommand{\empparm}{\widehat{\parm}}
\newcommand{\tgtpdf}{p_{\textsc{los}}}

\newcommand{\sseq}{s}
\newcommand{\seq}{\mathcal{S}}
\newcommand{\obspdf}{p}
\newcommand{\corrpdf}{p_{\textsc{nlos}}}

\newcommand{\prb}{\pi}
\newcommand{\prbvec}{\mbs{\pi}}
\newcommand{\entrp}{\mathbb{H}}

\newcommand{\prbset}{\Pi}
\newcommand{\intr}{\mu}
\newcommand{\intrvec}{\mbs{\intr}}
\newcommand{\best}{\parm^*}

\newcommand{\outlier}{\mathcal{C}}

\DeclareMathOperator{\E}{\mathbb{E}}

\DeclareMathOperator*{\argmin}{arg\,min}

\title{Robust Localization in Wireless Networks From Corrupted Signals}

\author{Muhammad Osama\thanks{Division of System and Control, Department of Information Technology, Uppsala University, Sweden. muhammad.osama@it.uu.se, dave.zachariah@it.uu.se. 
This work was supported by the Swedish Research Council, Research Environment NewLEADS (contract 2016-06079) and projects (contracts 2017-04610 and 2018-05040). }\\
	\And
	Dave Zachariah$^*$\\
	\AND
	Satyam Dwivedi\\
	\And
	Petre Stoica$^*$\\
}

\date{}

\begin{document}
\maketitle

\begin{abstract}
 We address the problem of timing-based localization in wireless networks, when an unknown fraction of data is corrupted by nonideal propagation conditions. While timing-based techniques can enable accurate localization, they are  sensitive to corrupted data. We develop a robust method that is applicable to a range of localization techniques, including time-of-arrival, time-difference-of-arrival and time-difference in schedule-based transmissions. The method is distribution-free and requires only an upper bound on the fraction of corrupted data, thus obviating distributional assumptions on the corrupting noise. The robustness of the method is demonstrated in numerical experiments.
\end{abstract}

\section{Introduction}

Localization in wireless networks is important for applications in GPS-denied environments \cite{caffery1998overview}.   The next generation wireless communication systems will standardize radio signals and measurements for localization in applications that  range from mobile broadband to industrial internet-of-things networks.
For accurate localization, these applications depend on techniques that use the signal time of flight between a transmitter and a receiver \cite{stoica2006lecture, beck2008exact}. 

\begin{figure*}[t!]
    \centering
    \begin{subfigure}{0.45\linewidth}
    \includegraphics[width=0.9\linewidth]{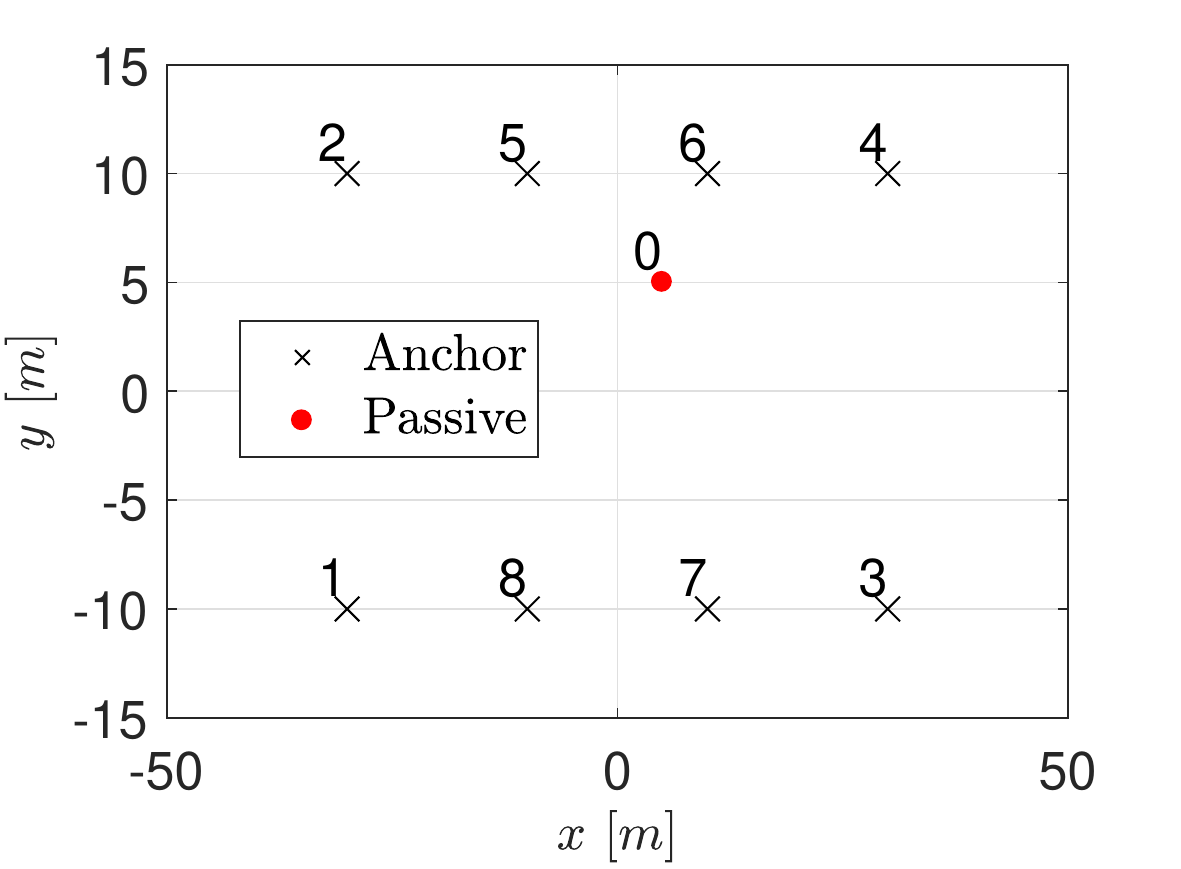}
    \caption{Wireless network}
    \label{fig:3gpp all nodes}
    \end{subfigure}
    \begin{subfigure}{0.45\linewidth}
    \includegraphics[width=0.9\linewidth]{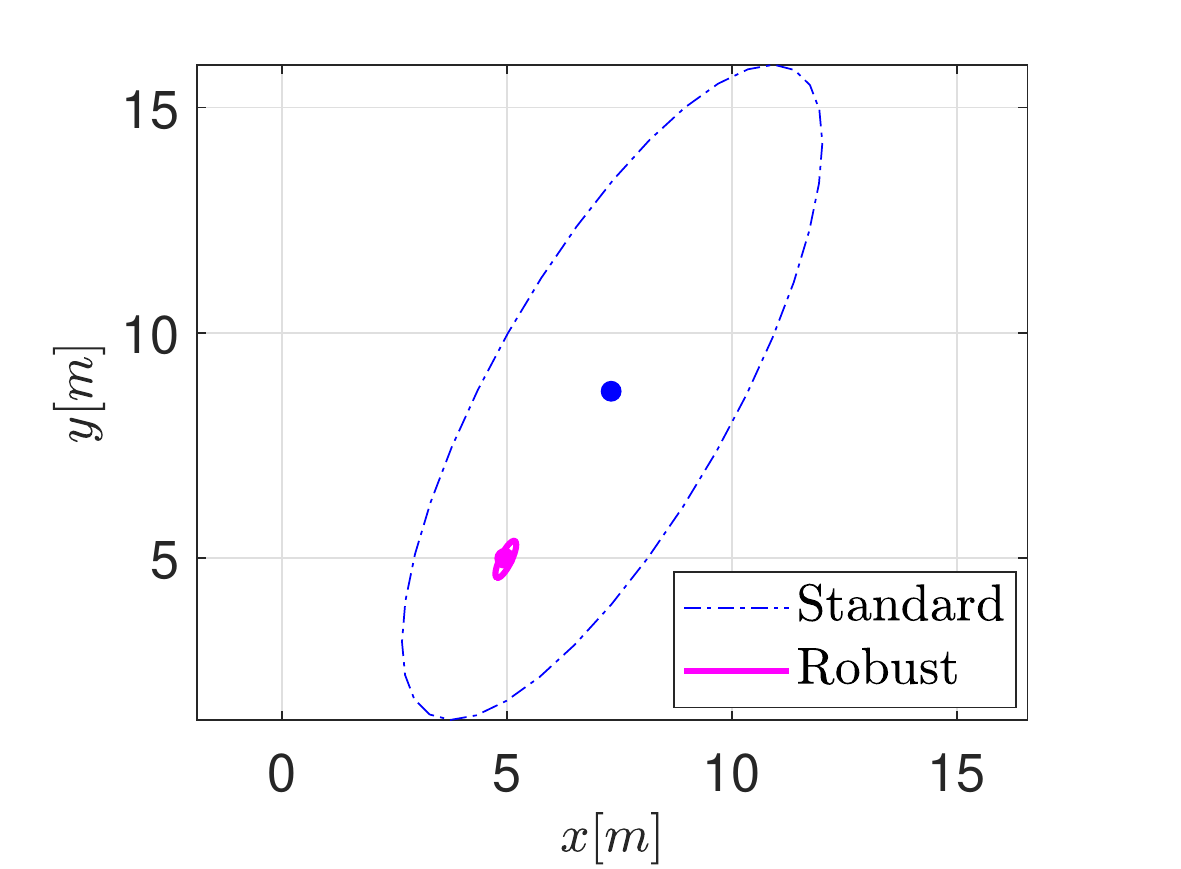}
    \caption{Estimates and their dispersions}
    \label{fig:3gpp tdoa}
    \end{subfigure}
    \caption{Self-localization of the passive receiving node ($\circ$) in (a) at location $(5,5)$ using eight anchor nodes ($\large{\times}$) and time-difference of arrival (\tdoa{}). The resolution of the interarrival times is on the order of $3 \text{ns}$ and $\Nobs=100$ observations are used. Here $\corr=15\%$ of the data is corrupted by non-line of sight (\nlos) outliers. In (b), \textcolor{blue}{\textbullet} and \textcolor{magenta}{\textbullet} denote the mean position estimated using the standard nonlinear least-squares method \eqref{eq:erm} and the proposed robust method. The ellipses illustrate the dispersion of location estimates with  approximately $99.9\%$ coverage for the standard (dotted) and robust (solid) methods.}  
    \label{fig:3gpp}
\end{figure*}

Real-world measurements of wireless signals are prone to outliers which arise not only from sensor failures but also from signals that fail to reach the receiver in an ideal line-of-sight (\los) manner. That is, wireless signals that may arrive at the receiver after reflections, diffractions or penetrations of different media, which we refer to as non-line-of-sight (\nlos) situations. The deployment of wireless technologies faces signal obstructions, moving objects, reflecting paths, foliage, etc. Under such non-ideal and \nlos{} conditions, standard measurement noise assumptions are invalid and conventional localization methods break down, as illustrated in Figure~\ref{fig:3gpp}.

Several approaches to robust localization exist. A conservative approach is to model the \nlos{} effects as bounded additive measurement errors or bias \cite{shi2016robust, tomic2017robust} and then estimate the node location that minimizes the worst-case error. This approach, however, can be overly conservative and sensitive to the user specified error bound. Another approach from the robust statistics literature is to consider the Huber contamination model \cite{huber2011robust}, in which an $\corr$ fraction of data samples are corrupted by a \nlos{}-data distribution. The \nlos{} distribution is either assumed to have a specific parametric form -- e.g. shifted Gaussian or exponential \cite{yin2013robust,mcguire2005data} in which case the positions of nodes are estimated via maximum likelihood -- or is modelled nonparametrically \cite{yin2013toa,hammes2008semi,hammes2009robust}, a case which is tackled using semiparametric or iterative maximum likelihood methods. An alternative approach is to use a different, robust loss function that is insensitive to outliers \cite{zoubir2018robust}. While it is often assumed that the fraction of corrupted data is known, in practice this fraction is an unknown user parameter.

In this paper, we propose a robust localization methodology for data obtained in the contamination setting with an unknown fraction $\corr$ of \nlos{} data, drawing upon the principles of robust risk minimization in \cite{osama2020robust}. We only assume that the user is able to set an upper bound $\corrub \geq \corr$. We demonstrate the methodology for three distinct localization techniques: time-of-arrival (\toa), time-difference-of-arrival (\tdoa) and time-difference in schedule-based transmissions (\async).


\section{Problem formulation}
For a general problem formulation, consider a wireless network consisting of $N+1$ nodes at locations
\begin{equation*}
\{\loc_\rcv,\underbrace{\loc_1,\ldots,~\loc_{\Na}}_{\text{auxiliary}},\underbrace{\loc_{\Na+1},\ldots,~\loc_{\N}}_{\text{anchors}}\}
\end{equation*}
in $\R^{d}$ space, where $d=2$ or $d=3$. Only the anchor node locations are known. The anchors and auxiliary nodes all transmit signals that propagate through space with constant velocity. These nodes can be thought of as base stations with known and unknown locations, respectively. The signals carry signatures that enable the identification of the transmitting node.

The self-localizing node, located at $\loc_\rcv$, can either be an active transceiver or a passive receiver, depending on the localization technique considered below. We let $\best$ denote the set of locations of interest and our goal is to estimate this set using timing measurements at node $\rcv$. 


\begin{figure*}[t!]
    \centering
    \begin{subfigure}{0.3\columnwidth}
    \includegraphics[width=1.0\linewidth]{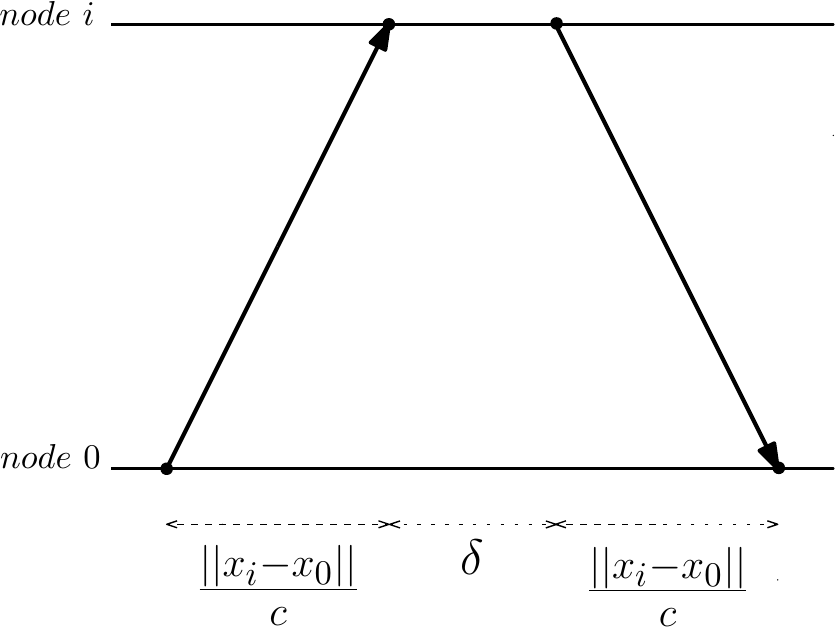}
    \caption{\toa}
    \label{fig: timing toa}
    \end{subfigure}
     \begin{subfigure}{0.3\columnwidth}
    \includegraphics[width=1.0\linewidth]{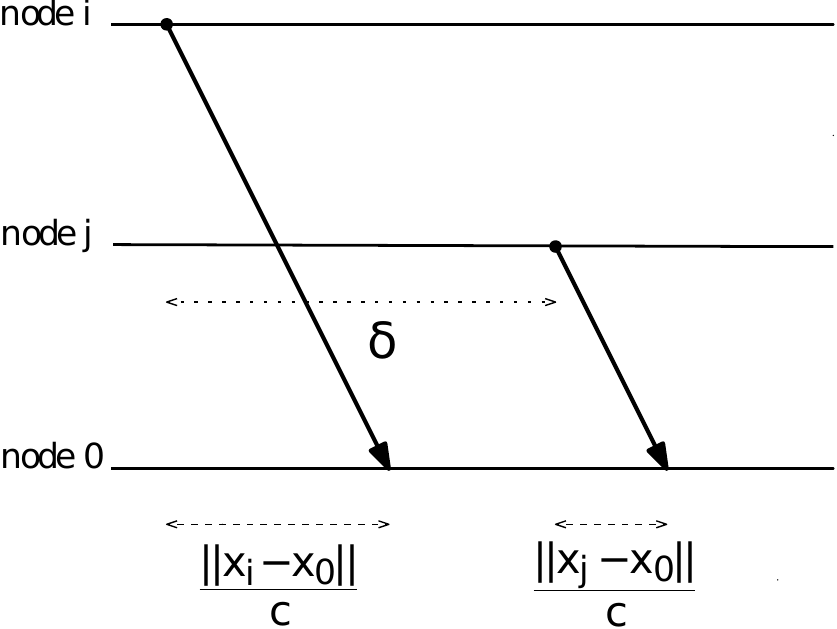}
    \caption{\tdoa}
    \label{fig: timing tdoa}
    \end{subfigure}
    \begin{subfigure}{0.3\columnwidth}
    \includegraphics[width=1.0\linewidth]{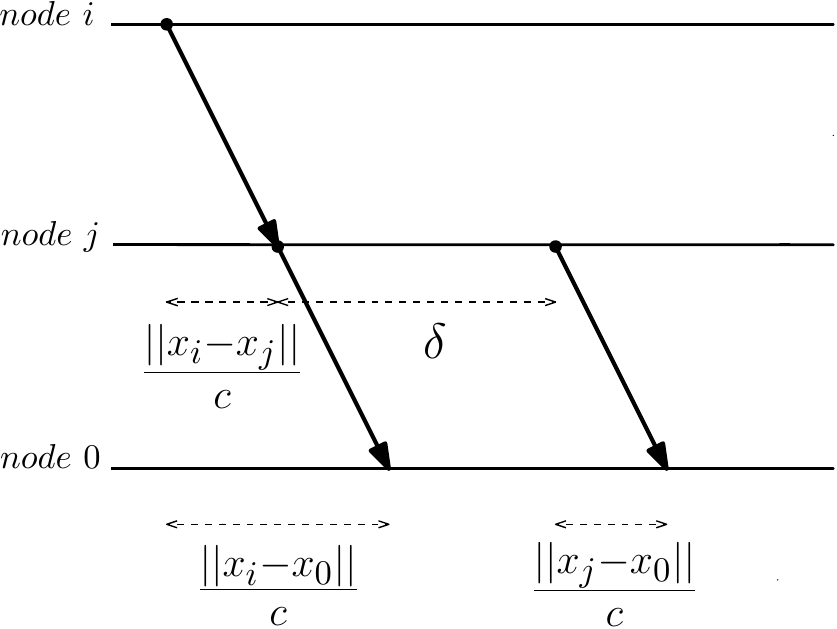}
    \caption{\async}
    \label{fig: timing tdst}
    \end{subfigure}
    \caption{Timing diagrams illustrating node transmitting signals in three different localization techniques.} 
    \label{fig:timing diag}
\end{figure*} 


\subsection{\toa{}: two-way ranging anchor nodes}
In \toa, node $\loc_\rcv$ is a transceiving node. At a given time, node $\rcv$ initiates a transmission to the anchor nodes. Upon receiving the signals, anchor nodes return signals, after a possible fixed delay. This leads to time-of-arrival measurements at self-localizing node $\rcv$. The $\toa$ measurement with respect to anchor node $i$ is then    
\begin{equation} \label{eq: time arrival scalar toa}
    \intr^{i0}= 2\frac{||\loc_{i}-\loc_{\rcv}||}{c} + \delay,
\end{equation}
where $c$ is the signal propagation velocity and $\delay\geq0$ is a transmission delay, see Figure~\ref{fig: timing toa}. The unknown location is 
$$\best~=~\{\loc_\rcv\}$$

\subsection{\tdoa{}: synchronous anchor nodes}

In \tdoa{}, the anchor nodes are synchronized so that they broadcast signals with respect to a common clock. Consider a pair of anchor nodes $(i,j)$ transmitting signals. Their interarrival time at the passive self-localizing receiver node $\rcv$ is then
\begin{equation} \label{eq: inter arrival scalar tdoa}
    \intr^{ij}= \frac{||\loc_{j}-\loc_{\rcv}||}{c} + \delay - \frac{||\loc_{i}-\loc_{\rcv}||}{c},
\end{equation}
where $\delay \geq 0$ is a (possible) transmission delay at node $i$, see Figure~\ref{fig: timing tdoa}. The unknown location  is
\begin{equation*}
\best~=~ \{ \loc_\rcv \} 
\end{equation*}

\subsection{\async{}: asynchronous anchor nodes}

In \async, the anchor nodes operate asynchronously so that the transmitted signals are only coordinated through a sequence of observable signal events. Consider a pair of transceiving nodes $(i,j)$, such that node $j$ transmits only \emph{after} receiving the signal from $i$ and a given delay. The interarrival time at the passive node is then
\begin{equation} \label{eq: inter arrival scalar schedule}
    \intr^{ij}=\frac{||\loc_{i}-\loc_{j}||}{c} + \frac{||\loc_{j}-\loc_{\rcv}||}{c} +\delay- \frac{||\loc_{i}-\loc_{\rcv}||}{c}. 
\end{equation}
where $\delay > 0$ is the transmission delay at node $j$, see Figure~\ref{fig: timing tdst}. As above, we consider
\begin{equation*}
\best~=~ \{ \loc_\rcv,  \}
\end{equation*}
However, we can readily accommodate unknown locations of auxiliary transmitting nodes in $\best$ as demonstrated in the experimental section. 

\subsection{Measurements in \los{} and \nlos{}}

We let $\sseq \subseteq \{ 1,2, \dots, N \}$ denote a set of transmitting nodes whose signals are observed by the self-localizing node. For a given $\sseq$, the ideal interarrival times, from either \eqref{eq: time arrival scalar toa}, \eqref{eq: inter arrival scalar tdoa} or \eqref{eq: inter arrival scalar schedule}, can be arranged into a vector expressed as
\begin{equation}\label{eq:meantimes}
\boxed{\intrvec(\sseq,\best)=\frac{1}{c}\M(\sseq)\rhovec(\best) + \delay \mbf{1},}
\end{equation}
where 
\begin{equation*}
\begin{split}
\rhovec(\best)=\begin{bmatrix}
\|\loc_1-\loc_2\| \\
\|\loc_1-\loc_3\| \\
\vdots \\
\|\loc_1-\loc_N\| \\
\|\loc_1-\loc_0\| \\
\vdots \\
\|\loc_N-\loc_0\|
\end{bmatrix}
\end{split}
\end{equation*}
is an $\frac{N(N+1)}{2} \times 1$ vector of distances between all nodes, $\M(\sseq)$ is a selection matrix of integers and $\delay$ is a known delay parameter.

\begin{exmp}[\toa{}] \label{ex:toa}
For $\N=3$ and $\sseq=\{1,2,3\}$, we have $ \intrvec(\sseq,\best)~=~[\intr^{10},~\intr^{20}, ~\intr^{30}]^\T$ and 
\begin{equation*}
\M(\sseq)~=~
\begin{bmatrix}
0 & 0 & 2 & 0 & 0 & 0\\
0 & 0 & 0 & 0 & 2 & 0\\
0 & 0 & 0 & 0 & 0 & 2
\end{bmatrix}
\end{equation*}
\end{exmp}
When $\sseq$ contains at least three anchor nodes that are not coaligned, $\best$ is uniquely determined from the set of interarrival times, see \cite{ravindra2014time}.

\begin{exmp}[\tdoa] \label{ex:tdoa}
For $\N=3$ and $\sseq=\{1,2,3\}$, we have $ \intrvec(\sseq,\best)~=~[\intr^{12},~\intr^{23}]^\T$ and
\begin{equation*}
\M(\sseq)~=~
\begin{bmatrix}
0 & 0 & -1 & 0 & 1 & 0\\
0 & 0 & 0 & 0 & -1 & 1
\end{bmatrix}
\end{equation*}
in \eqref{eq:meantimes}.
\end{exmp}
When $\sseq$ contains at least three anchor nodes that are not coaligned, $\best$ is uniquely determined from the set of interarrival times, see \cite{gustafsson2003positioning}.

\begin{exmp}[\async]
For $\N=3$ and $\sseq=\{1,2,3,1\}$, we have $ \intrvec(\sseq,\best)~=~[\intr^{12},~\intr^{23},~\intr^{31}]^\T$ and 
\begin{equation*}
\M(\sseq)~=~
\begin{bmatrix}
1 & 0 & -1 & 0 & 1 & 0\\
0 & 0 & 0 & 1 & -1 & 1\\
0 & 1 & 1 & 0 & 0 & -1
\end{bmatrix}
\end{equation*}
in \eqref{eq:meantimes}.
\end{exmp}
When $\sseq$ contains at least three anchor nodes that are not coaligned, $\best$ can be uniquely determined from the set of interarrival times, see  \cite[sec.~3.3]{zachariah2014schedule} and \cite{dwivedi2012scheduled,cavarec2017schedule}.

In ideal \los{} conditions, a set is drawn as $\sseq \sim \tgtpdf(\sseq)$, where the distribution is possibly degenerate. For each observed set $s$, the self-localizing node obtains noisy measurements $\obsvec$ of the interarrival times $\intrvec(\sseq, \best)$. That is,
\begin{equation}
\obsvec \sim \tgtpdf(\obsvec|\sseq)
\end{equation}
where the mean of $\tgtpdf(\obsvec|\sseq)$ equals $\intrvec(\sseq, \best)$ in \eqref{eq:meantimes}.

In practical wireless environments, however, the pair $(\sseq, \obsvec)$ is not always observed in ideal conditions, but rather drawn from
\begin{equation} \label{eq:obspdf}
\boxed{\obspdf(\sseq,\obsvec)=(1-\corr)~\tgtpdf(\sseq,\obsvec) + \corr~\corrpdf(\sseq,\obsvec),}
\end{equation}
where $\corr$ is an unknown fraction of corrupted data and $\corrpdf(\sseq, \obsvec)$ is an unknown corrupting distribution, that generates outlier noise and biases such that its mean may differ from $\intrvec(\sseq, \best)$ \cite{huber2011robust,zoubir2018robust}.
The problem under consideration is to estimate $\best$ given data $\{(\sseq_1,\obsvec_1), \dots,(\sseq_n,\obsvec_n) \}$ drawn i.i.d. from \eqref{eq:obspdf}. We merely assume that the quality of the data can be specified by an upper bound $\corrub \geq \corr$ \cite[ch.~1]{hampel2011robust}.

\section{Method}
Assuming a set of anchor locations and transmission sequences that yield identifiable locations, we have that 
\begin{equation}\label{eq:best model}
    \best\equiv\argmin_{\parm}~\E\Big[\| \obsvec-\intrvec(\sseq,\parm)\|^2\Big],
\end{equation}
where the expectation is with respect to the unknown distribution $\tgtpdf(\sseq, \obsvec)$. Note that the right-hand side of \eqref{eq:best model} assumes that $\obsvec$ has finite second-order moments. The standard estimator of $\best$ is the nonlinear least-squares method,
\begin{equation}\label{eq:erm}
    \empparm=\argmin_{\parm}~\frac{1}{n}\sum_{i=1}^{\Nobs} \| \obsvec_i-\intrvec(\sseq_i,\parm)\|^2 ,
\end{equation}
where $(\sseq_i, \obsvec_i)\sim p(\sseq, \obsvec)$ in \eqref{eq:obspdf}. Under standard regularity  conditions $\empparm$ is consistent when $\corr = 0$, and corresponds to the maximum likelihood estimate assuming white Gaussian noise \cite{Kay1993_sspestimation}. For $\corr > 0$, however, $\empparm$ is not robust to corrupted samples that arise in \nlos{} conditions, as described by \eqref{eq:obspdf}. 

\subsection{Robust localization}  
As an alternative to \eqref{eq:best model}, consider the following fitting criterion
\begin{equation}\label{eq:weighted loss}
    \E_{\prbvec}\Big[\| \obsvec-\intrvec(\sseq,\parm)\|^2\Big],
\end{equation}
where, in lieu of $\tgtpdf(\sseq, \obsvec)$ in \eqref{eq:best model}, we use the distribution
\begin{equation}
    \obspdf(\sseq,\obsvec; \prbvec)=\sum_{i=1}^{\Nobs}\prb_{i}\delta_{\sseq_i,\obsvec_i }(\sseq,\obsvec),
\end{equation}
with probability weights $\prbvec = [\pi_1, \dots, \pi_n]^\T \in \prbset$, where $\prbset$ is the probability simplex. This distribution has an entropy
\begin{equation}
    \entrp(\prbvec)\triangleq-\sum_{i=1}^{\Nobs}\prb_i\ln\prb_i~\geq~0
\end{equation}
It follows that the standard method \eqref{eq:erm} corresponds to minimizing \eqref{eq:weighted loss} using the empirical distribution $\obspdf(\sseq,\obsvec; n^{-1} \mbf{1})$, which attains a maximum entropy of $\ln n$. 

Given the bound $\corrub \geq \corr$, however, we expect at least 
$(1-\corrub)\Nobs$ uncorrupted samples and that the support of $\obspdf(\sseq,\obsvec; \prbvec)$ should cover them. In this case, the maximum entropy would equal $\ln[(1-\corrub)\Nobs]$ and the search over the unknown support turns \eqref{eq:weighted loss} into the following joint optimization problem
\begin{equation}\label{eq:main optimization problem}
    \begin{split}
       \min_{\parm,~\prbvec \in \prbset}~&\sum_{i=1}^{\Nobs}\prb_i \| \obsvec_i-\intrvec(\sseq_i,\parm)\|^2\\
       \text{subject to}~~\entrp(\prbvec)&\geq\ln\big[(~1-\corrub~)\Nobs\big]
    \end{split}
\end{equation}
Intuitively, the above minimization problem estimates $\parm$ and simultaneously assigns weights $\prb_i$ to each point such that the overall weighted squared-error loss is minimized under the entropy constraint. Smaller weights are assigned to datapoints which are corrupted due to \nlos~and larger weights are assigned to uncorrupted points. In this way, \eqref{eq:main optimization problem} enables robust localization of the auxiliary and receiver nodes without distributional assumptions. In the next subsection, we give a blockwise algorithm for solving \eqref{eq:main optimization problem}.

\subsection{Blockwise minimization
 algorithm}
For a fixed $\widetilde{\parm}$, define
\begin{equation} \label{eq:opt prb}
    \widehat{\prbvec}(\widetilde{\parm})=
    \begin{cases}
    \underset{\prbvec\in\prbset}{\text{arg~min}}~\sum_{i=1}^{\Nobs}\prb_i \| \obsvec_i-\intrvec(\sseq_i,\parm)\|^2,\\
   \text{s.t.}~\entrp(\prbvec)\geq \ln\big[(1-\corrub)\Nobs\big]
    \end{cases}
\end{equation}
and, similarly, for a fixed $\widetilde{\prbvec}$, define
\begin{equation} \label{eq:opt theta}
    \empparm(\widetilde{\prbvec})=\argmin_{\parm}~\sum_{i=1}^{\Nobs} \widetilde{\prb}_i\| \obsvec_i-\intrvec(\sseq_i,\parm)\|^2
\end{equation}
Together \eqref{eq:opt prb} and \eqref{eq:opt theta} form a blockwise coordinate descent algorithm which we summarize in Algorithm~\ref{algo:robust local}, that is guaranteed to converge to a critical point of \eqref{eq:main optimization problem} under fairly general conditions \cite{grippo2000convergence}. The weighted nonlinear least-squares problem in \eqref{eq:opt theta} can be solved using standard search methods (e.g. gradient-based or Newton search), while \eqref{eq:opt prb} requires solving a convex problem, whose solution can be obtained using a barrier method that is more efficient than general purpose numerical packages such as \texttt{cvx} \cite{boyd2004convex}.
\begin{algorithm}
\caption{Robust Localization} \label{algo:robust local}
\begin{algorithmic}[1]
\STATE Input: $\{(\sseq_i,\obsvec_i)\}_{i=1}^{\Nobs}$, $\corrub$ and $\prbvec^{(0)}~=~n^{-1}\mbf{1}$
\STATE Set $k~:=~0$
\STATE \textbf{repeat}
\STATE $\parm^{(k+1)}~=~\widehat{\parm}(\prbvec^{(k)})$
\STATE $\prbvec^{(k+1)}~=~\widehat{\prbvec}(\parm^{(k+1)})$
\STATE $k~:=~k~+~1$
\STATE \textbf{until convergence}
\STATE Output: $\widehat{\parm}~=~\parm^{(k)}$,~$\widehat{\prbvec}~=~\prbvec^{(k)}$
\end{algorithmic}
\end{algorithm}\\

\section{Experimental results}
In this section, we illustrate the wide applicability of the proposed robust localization method using synthetic \toa{}, \tdoa{} and \async{} data. The performance is evaluated using the localization error
$$\Delta(\widehat{\loc})~=~||\loc-\widehat{\loc}||$$
where $\loc$ is the node location of interest.

We observe $\Nobs~=~100$ measurements from \eqref{eq:obspdf}. The \los{} distribution is
\begin{equation} \label{eq:exp target pdf}
    \tgtpdf(\sseq,\obsvec)=\underbrace{\mathcal{N}\left(\intrvec(\sseq,\best),\sigma_{\los}^2 \mbf{Q} \right)}_{\tgtpdf(\obsvec|\sseq)}~\underbrace{\mathcal{U}(\sseq)}_{\tgtpdf(\sseq)},
\end{equation}
where adjacent timing measurement errors have correlation structure given by $\mbf{Q}$. The uniform distribution $\mathcal{U}(\sseq)$ draws $\sseq$ from a set $\seq$. 
The corrupting \nlos{} distribution is 
\begin{equation} \label{eq:exp corr pdf}
    \corrpdf(\sseq,\obsvec)=\underbrace{\mathcal{E}(\intrvec(\sseq,\best)+\mu_{\nlos})}_{\corrpdf(\obsvec|\sseq)}~\underbrace{\mathcal{U}(\sseq)}_{\corrpdf(\sseq)},
\end{equation}
where we consider an exponential distribution with measurement bias $\mu_{\nlos}$. We set $\sigma_{\los}~=~3~\text{ns}$ and  $\mu_{\nlos}~=~75~\text{ns}$. Unless otherwise specified, the unknown corruption fraction is set to $\corr = 15\%$.

\emph{Remark:} The code for the experiments is available at \href{https://github.com/Muhammad-Osama/RobustLocalization}{github.com/Muhammad-Osama/RobustLocalization}.

\subsection{\toa{}: two-way ranging anchor nodes}\label{sec:two way ranging}
Consider a wireless network consisting of $\N~=~8$ nodes as shown in Figure \ref{fig:3gpp all nodes}. Since the \toa{} measurements are uncorrelated, $\mbf{Q}= \mbf{I}$. The unknown location of interest is
$$\best~=~\{\loc_{0} \} = \left\{ [5,~5]^\T \right\}.$$
The set $\seq$ consists of two sequences $\sseq_{0}~=~\{6,5,7,8\}$ and $\sseq_{1}~=~\{4,3,2,1\}$ where the node numbers  are given in Figure \ref{fig:3gpp all nodes}. The sequences have been selected so that the anchor nodes are not coaligned in either $\sseq_0$ or $\sseq_1$.

In the case of \toa, the measurement model \eqref{eq: time arrival scalar toa} admits an (overparameterized) linear form which is ideal for classical methods in robust statistics, such as the Huber method \cite{zoubir2018robust}. The Huber method is thus tailored for the task of \toa{} in which it provides a useful benchmark. We compare it to the standard nonlinear least-squares method \eqref{eq:erm} and the proposed method \eqref{eq:main optimization problem}. 

\begin{figure*}[t!]
    \centering
    \begin{subfigure}{0.45\linewidth}
    \includegraphics[width=0.9\linewidth]{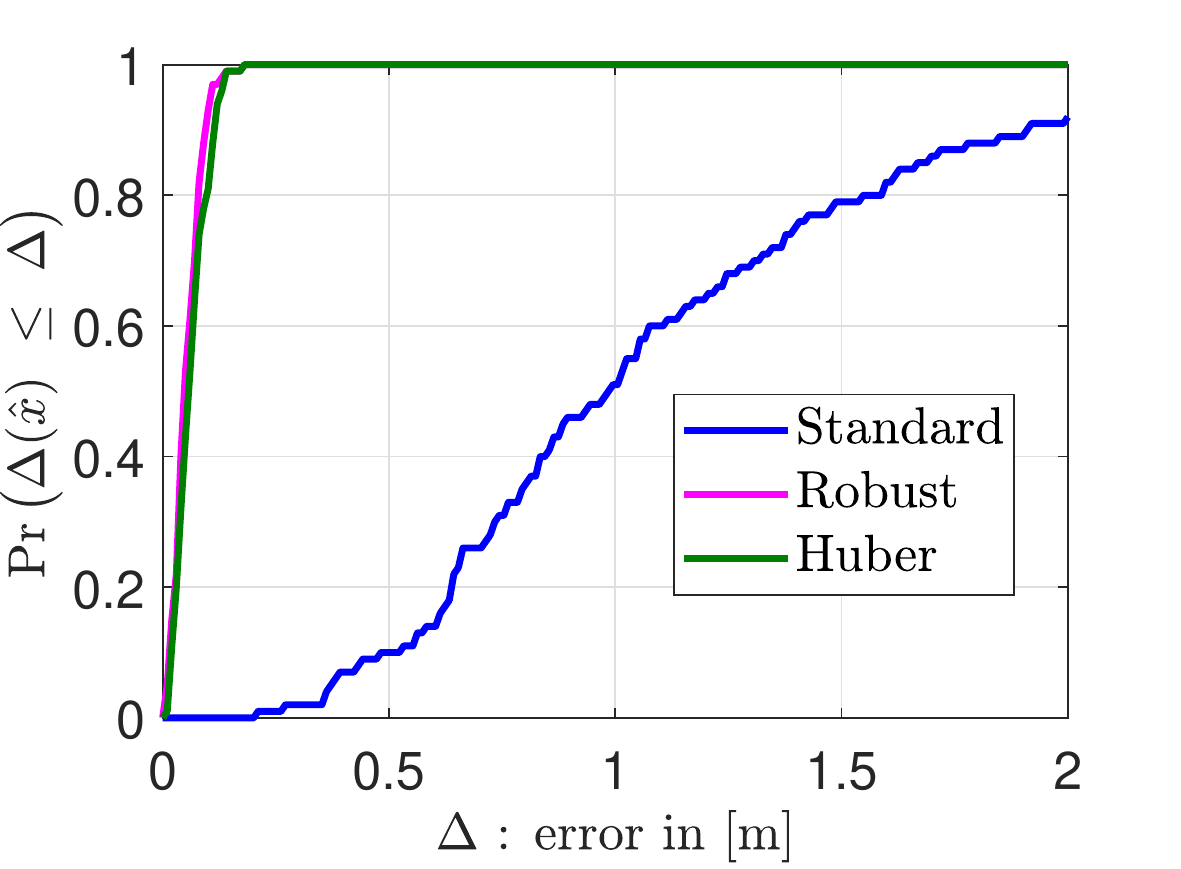}
    \caption{\cdf{}}
    \label{fig:toa ccdf}
    \end{subfigure}
    \begin{subfigure}{0.45\linewidth}
    \includegraphics[width=0.9\linewidth]{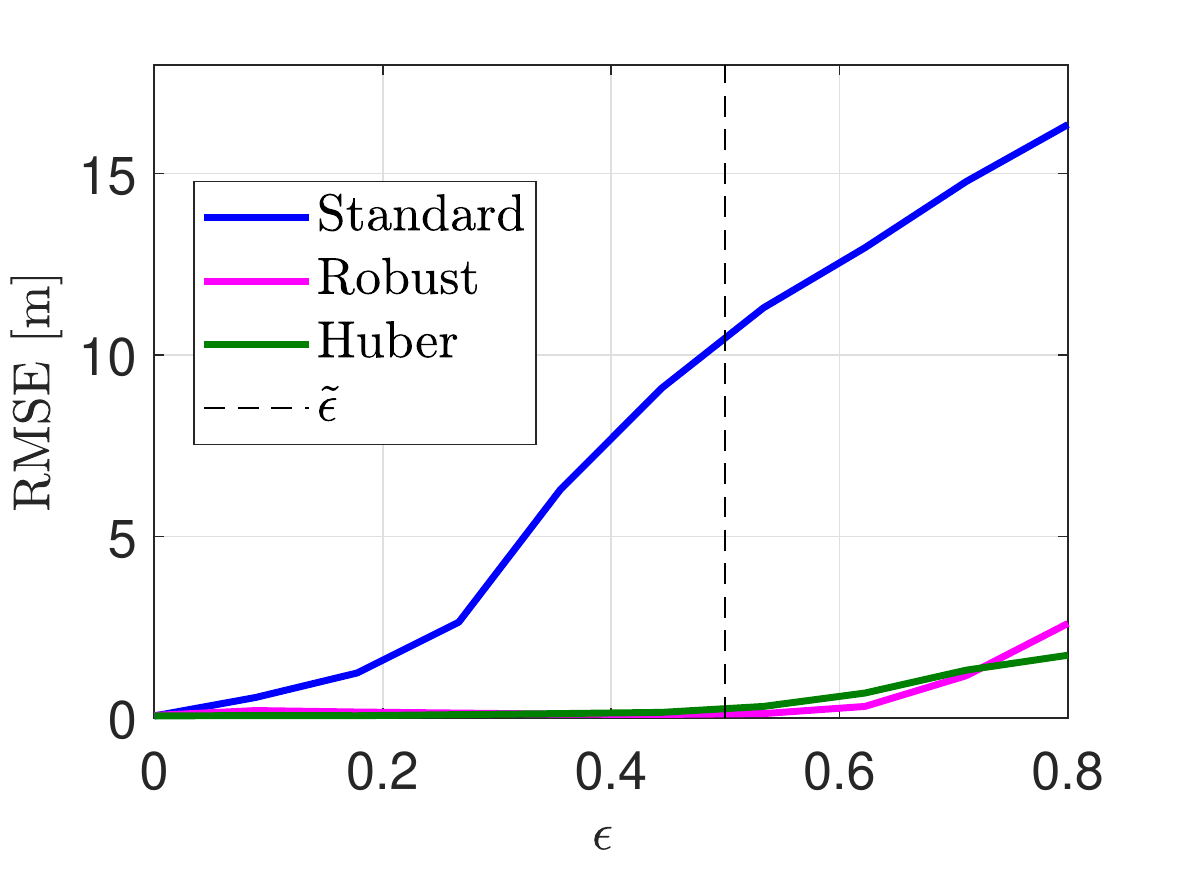}
    \caption{\rmse}
    \label{fig:toa rmse}
    \end{subfigure}
    \caption{Performance and sensitivity in \toa{}. (a) Cumulative distribution functions of localization errors $\Delta(\widehat{\loc})$ of target node in Figure \ref{fig:3gpp all nodes} using $100$ Monte Carlo runs. Unknown corruption fraction was $\corr~=~15\%$ and the upper bound used in the robust method was set to $\corrub~=~20\%$. (b) Root-mean square error in [m] as a function of $\corr$ for the target node using standard, robust and Huber methods. Results based on $50$ Monte Carlo simulations. For the proposed robust method the upper bound was $\corrub~=~50\%$.} 
    \label{fig:toa}
    \end{figure*}

Figure \ref{fig:toa ccdf} shows the cumulative distribution functions (\cdf{}) of the localization error $\Delta(\widehat{\loc})$ estimated from $100$ Monte Carlo simulations, setting $\corrub~=~20\%$ in Algorithm \ref{algo:robust local}. As expected there is a severe performance degradation of the standard method. To investigate the sensitivity of the results to the unknown corruption fraction, we also plot the root-mean-square error (\rmse), i.e. $\sqrt{\E[\Delta^{2}(\widehat{\loc})]}$,  versus $\corr$ for all three methods. For the proposed method we set a very conservative upper bound $\corrub~=~50\%$ in Algorithm \ref{algo:robust local}. Figure \ref{fig:toa rmse} shows the $\rmse$ for the different methods, where we see that the robust method is insensitive to $\corr$, with a graceful rise in \rmse{} when $\corr$ exceeds $\corrub$. In sum, the proposed method outperforms the standard nonlinear least-squares method and is close to the benchmark provided by the \toa{}-tailored Huber method. 

\begin{figure*}[h!]
    \centering
    \begin{subfigure}{0.3\linewidth}
    \includegraphics[width=0.9\linewidth]{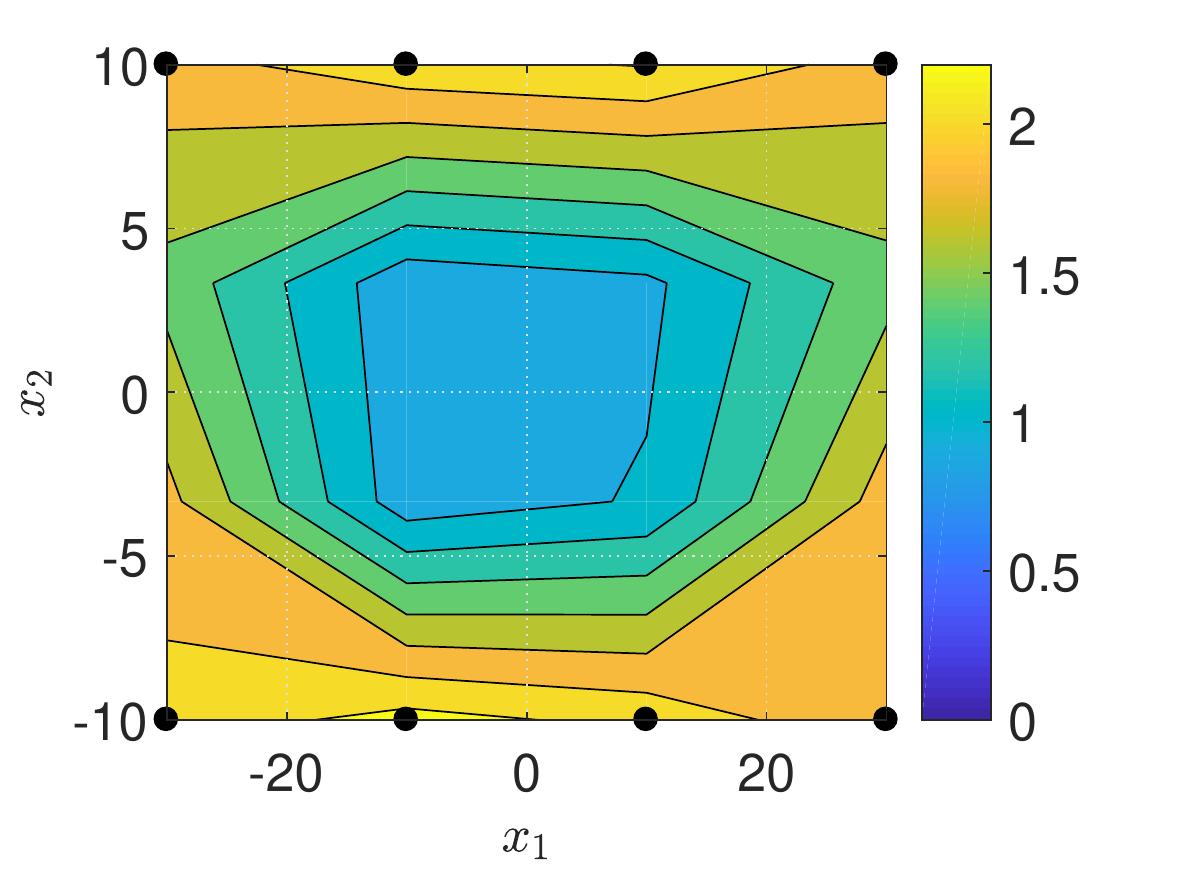}
    \caption{Standard method}
    \label{fig:toa cont stand}
    \end{subfigure}
    \begin{subfigure}{0.3\linewidth}
    \includegraphics[width=0.9\linewidth]{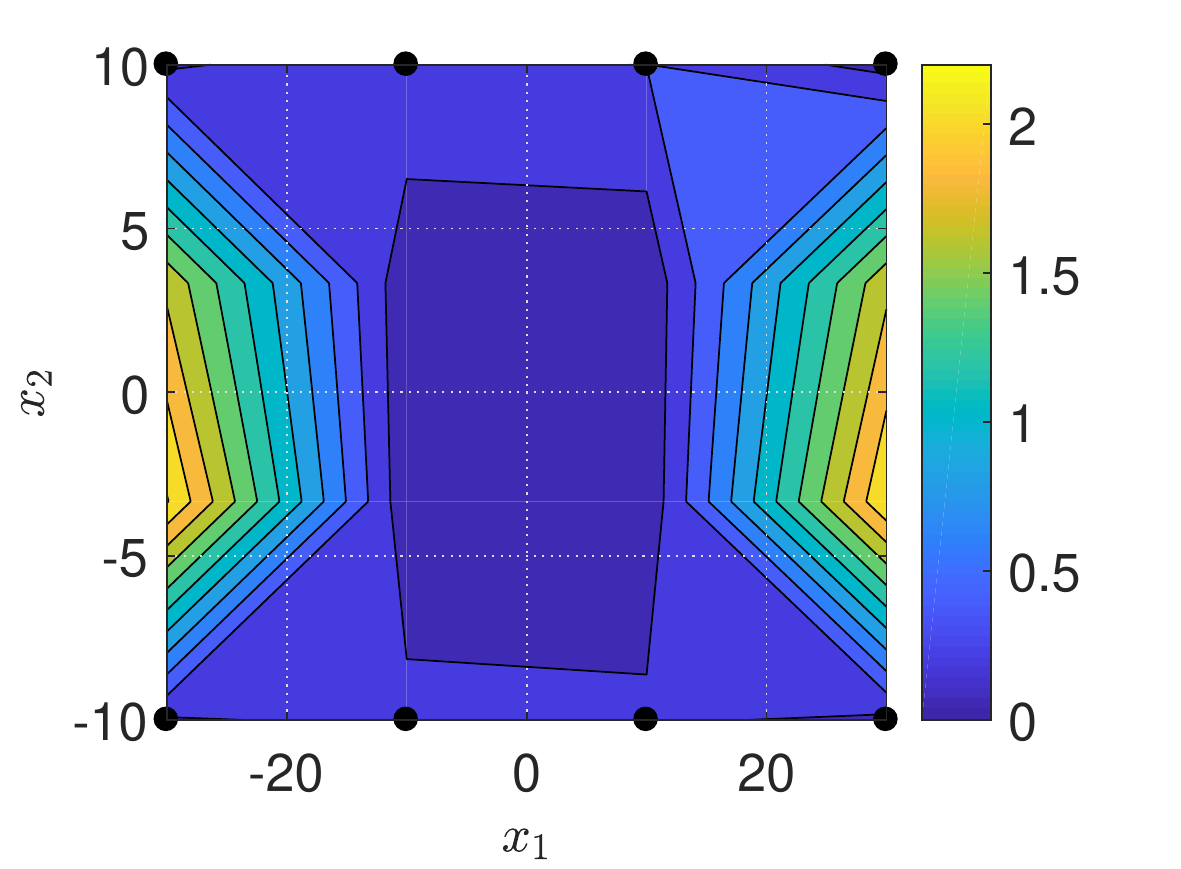}
    \caption{Robust method}
    \label{fig:toa cont rob}
    \end{subfigure}
    \begin{subfigure}{0.3\linewidth}
    \includegraphics[width=0.9\linewidth]{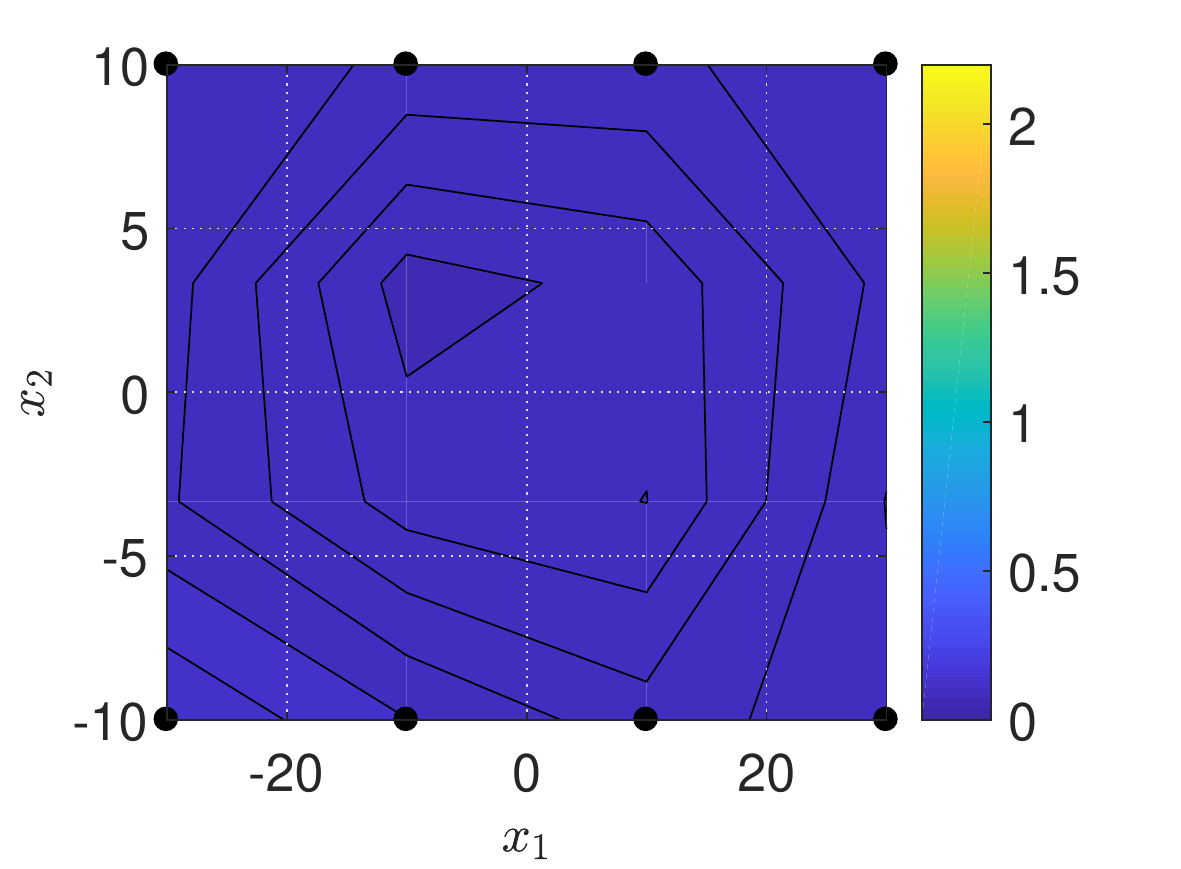}
    \caption{Huber method}
    \label{fig:toa cont hub}
    \end{subfigure}
    \caption{Performance in \toa{} across space.  Root-mean square error in [m] with respect to different locations $\loc_\rcv$. Results based on $50$ Monte Carlo runs.} 
    \label{fig:toa contour}
    \end{figure*}

The results are corroborated also in Figure \ref{fig:toa contour} which shows $\rmse$ as a function of $\loc_\rcv$. It can be seen that the robust method is  more sensitive than the benchmark only near the edges of the vertical boundaries, where the 

resolution of time-differences decreases.

\begin{figure*}[t!]
    \centering
    \begin{subfigure}{0.45\linewidth}
    \includegraphics[width=0.9\linewidth]{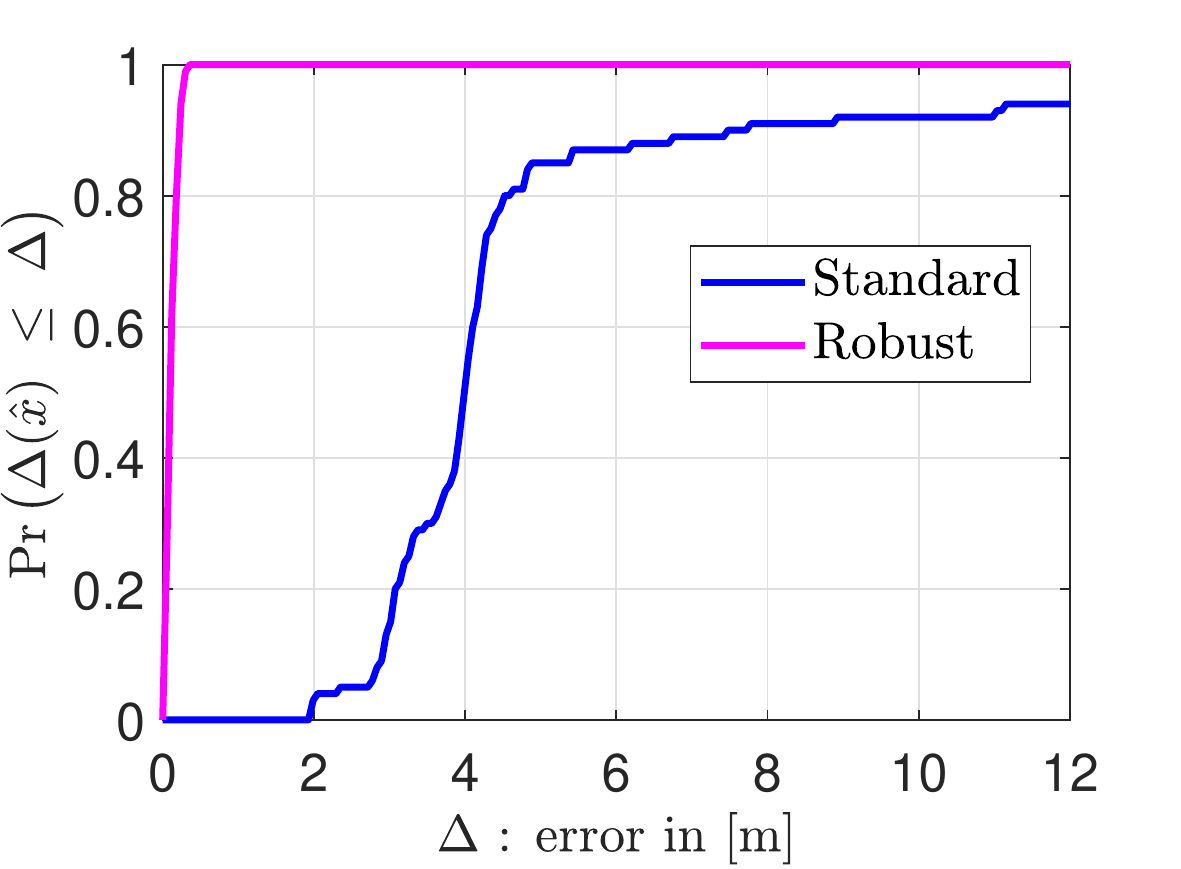}
    \caption{\cdf{}}
    \label{fig:TDOA_CDF}
    \end{subfigure}
    \begin{subfigure}{0.45\linewidth}
    \includegraphics[width=0.9\linewidth]{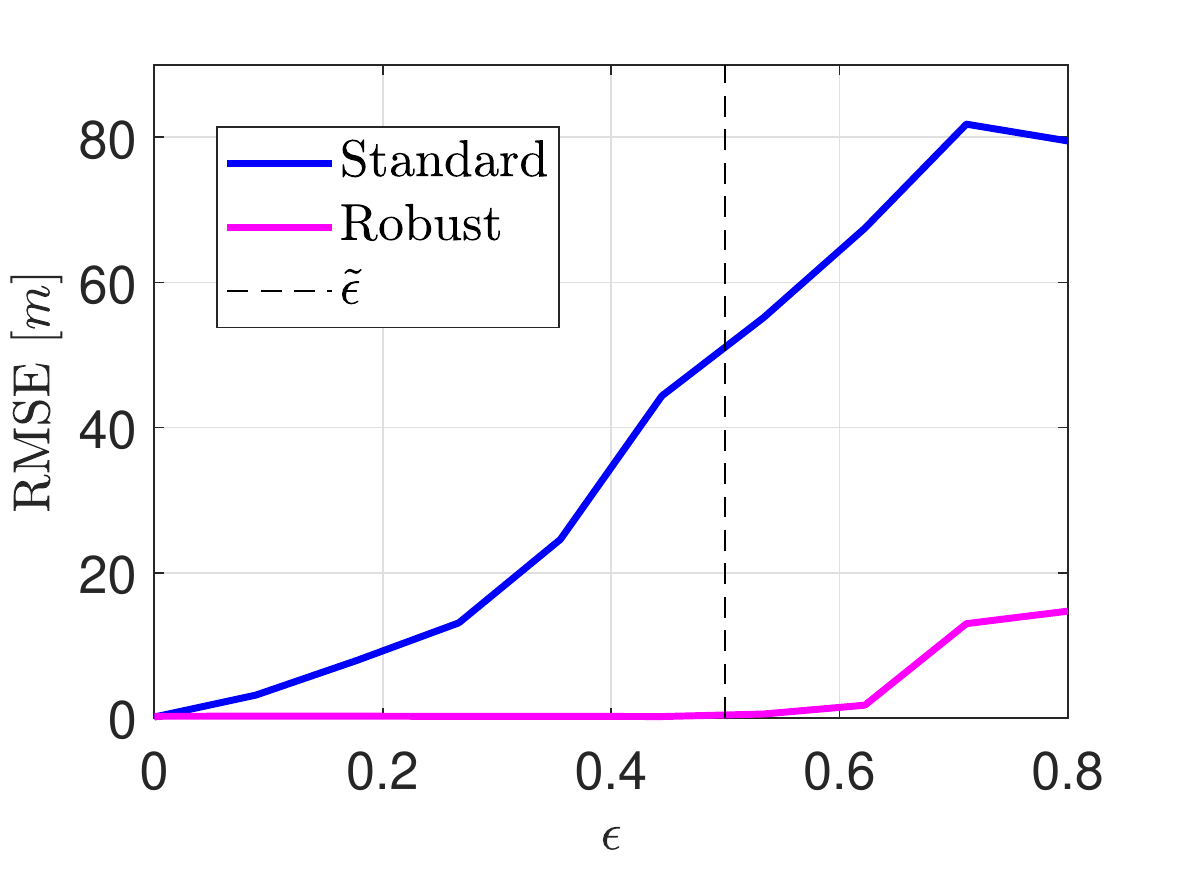}
    \caption{\rmse}
    \label{fig:TDOA_RMSE}
    \end{subfigure}
    \caption{Performance and sensitivity in \tdoa{}. (a) Cumulative distribution functions of localization errors $\Delta(\widehat{\loc})$ of target node in Figure \ref{fig:3gpp all nodes} using $100$ Monte Carlo runs. Unknown corruption fraction was $\corr~=~15\%$ and the upper bound used in the robust method was set to $\corrub~=~20\%$. (b) Root-mean square error in [m] as a function of $\corr$ for the target node using standard and robust methods. Results based on $50$ Monte Carlo simulations. For the proposed robust method the upper bound was $\corrub~=~50\%$.} 
    \label{fig:CDF}
    \end{figure*}

\begin{figure*}[h!]
    \centering
    \begin{subfigure}{0.45\linewidth}
    \includegraphics[width=0.9\linewidth]{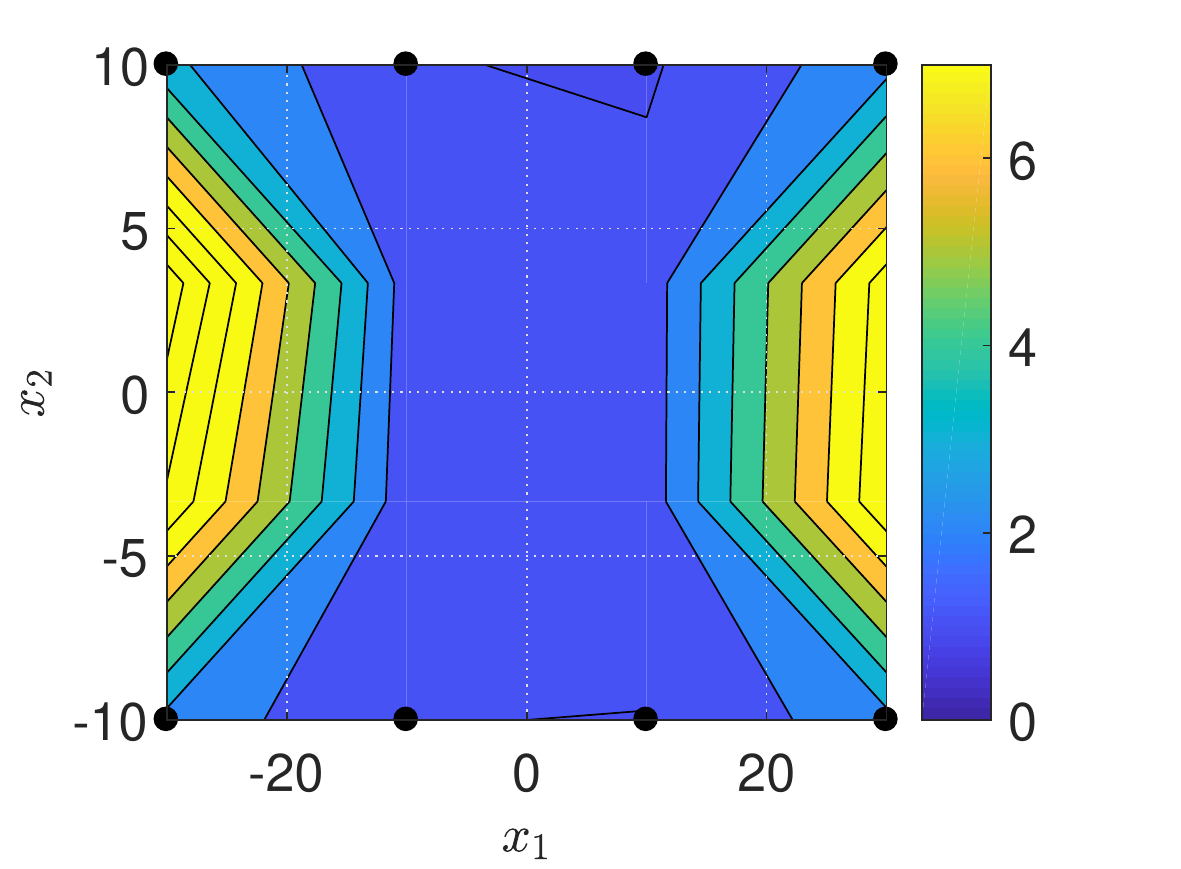}
    \caption{Standard method}
    \label{fig:TDOA_RMSE_STD}
    \end{subfigure}
    \begin{subfigure}{0.45\linewidth}
    \includegraphics[width=0.9\linewidth]{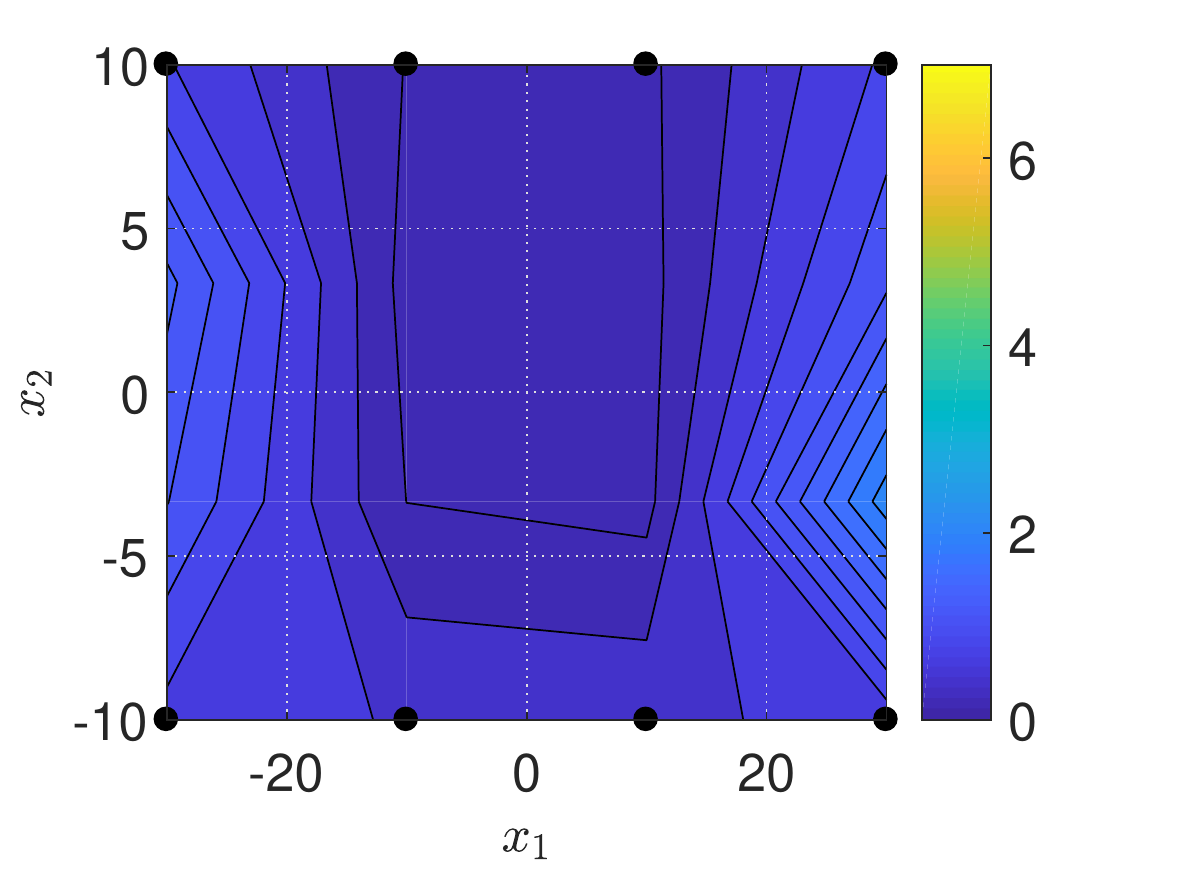}
    \caption{Robust method}
    \label{fig:TDOA_RMSE_ROB}
    \end{subfigure}
    \caption{Performance in \tdoa{} across space.  Root-mean square error in [m] with respect to different locations $\loc_\rcv$. Results based on $50$ Monte Carlo runs.}
    \label{fig:tdoa contour}
\end{figure*}

\subsection{\tdoa{}: synchronous anchor nodes}

We consider again the network in Figure \ref{fig:3gpp all nodes}, but now the self-localizing node is a passive receiver. Since \tdoa{} measurements are correlated, we set $\mbf{Q}$ with $1$s along the diagonal and $1/3$ along the off-diagonals. The unknown location of interest is
$$\best~=~\{\loc_{0} \} = \left\{ [5,~5]^\T \right\}.$$
As in the case of \toa{} above, we use the sequence set $\seq$  with $\sseq_{0}~=~\{6,5,7,8\}$ and $\sseq_{1}~=~\{4,3,2,1\}$.

Note that the measurement model \eqref{eq: inter arrival scalar tdoa} is nonlinear in $\parm$ and does not admit a linear re-parametrization. Thus the Huber method is not readily applicable for \tdoa{} as it requires tuning an alternative numerical search techniques. We therefore compare only the standard nonlinear least-squares method \eqref{eq:erm} and the proposed method \eqref{eq:main optimization problem}. 

Figure \ref{fig:TDOA_CDF} shows \cdf{}s of the localization error $\Delta(\widehat{\loc})$ estimated using $100$ Monte Carlo simulations, setting $\corrub~=~20\%$ in Algorithm \ref{algo:robust local}. The performance characteristics are similar to those in Figure~\ref{fig:toa ccdf}, but the absolute error levels are not directly comparable due to different measurement setups. The sensitivity to the unknown corruption fraction $\corr$ is also shown in Fig.~\ref{fig:TDOA_RMSE} when \rmse{} is plotted against $\corr$ for both methods. We use a very conservative upper bound $\corrub~=~50\%$ in Algorithm \ref{algo:robust local}. The proposed method is consistently robust and insensitive to corrupted data in contrast to the standard method for which the errors rise drastically with $\corr$. Figure \ref{fig:tdoa contour} shows that the robust method yields substantial error reduction across space.

\begin{figure}[t!]
    \centering
    \includegraphics[width=0.45\linewidth]{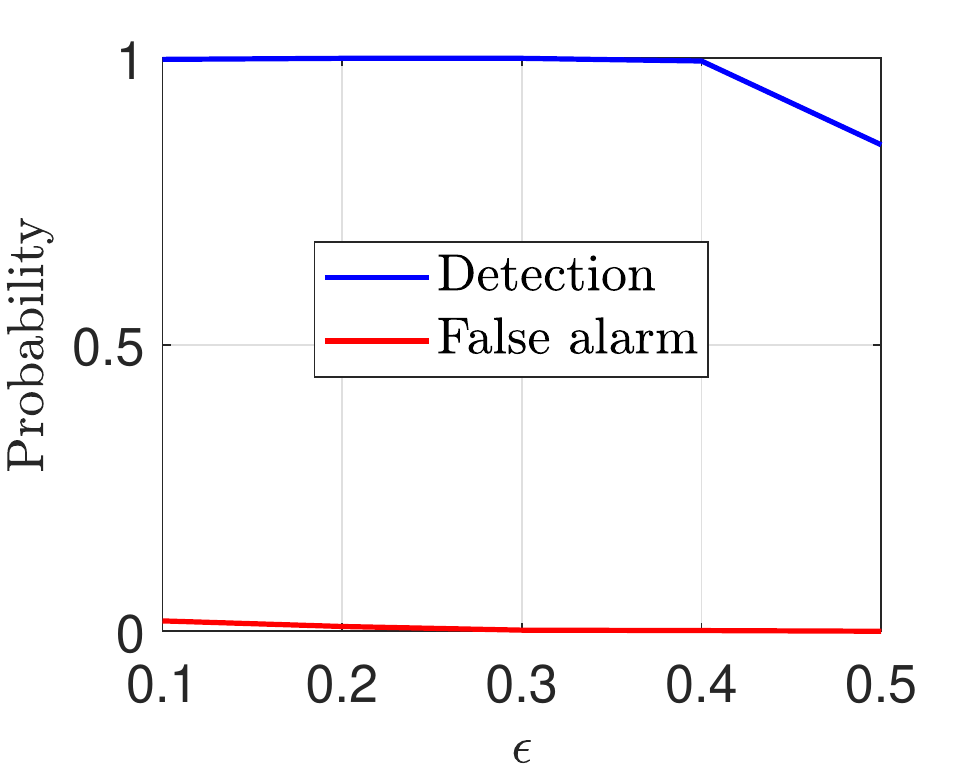}
    \caption{The weights in the robust method can be used to classify samples $\outlier(\obsvec) \in \{0,1 \}$ as `normal' or `corrupted'. This yields probabilities of correct detection $\Pr\{\widehat{\outlier}(\obsvec)=1|\outlier(\obsvec) = 1\}$ and false alarm $\Pr\{\widehat{\outlier}(\obsvec)=1|\outlier(\obsvec) = 0\}$ shown in the figure as the corruption fraction $\corr$ is varied from 10\% to 50\% $= \corrub$.} 
    \label{fig:nlos detection}
\end{figure}

Finally, we illustrate the ability of the proposed method to effectively isolate corrupted \nlos{} samples. Since the data is generated synthetically we can classify each sample from $\corrpdf(\obsvec|\sseq)$ as $\outlier(\obsvec)=1$ for `corrupted' or $\outlier(\obsvec)=0$ for `normal'. The method solves \eqref{eq:main optimization problem} and learns the probability weights $\widehat{\prbvec}$. If a weight is below a certain threshold, we may classify the corresponding sample $\obsvec$ as an outlier, i.e., $\widehat{\outlier}(\obsvec) = 1$. We set  the weight threshold to $10^{-5}$ and show the resulting probability of correct detection $\Pr\{\widehat{\outlier}(\obsvec)=1|\outlier(\obsvec) = 1\}$ as well as the probability of false alarm $\Pr\{\widehat{\outlier}(\obsvec)=1|\outlier(\obsvec) = 0\}$ in Figure~\ref{fig:nlos detection}. We use $\corrub = 50\%$ and vary the unknown fraction $\corr$, using $50$ Monte Carlo runs for each value of $\corr$.  It can be seen that the proposed method can  effectively isolate \nlos{} samples  with a low false-alarm rate. 

\subsection{\async{}: asynchronous anchor nodes}
We consider the network in Figure \ref{fig:3gpp all nodes}, but now the self-localizing node is a passive receiver and the anchor nodes are asynchronous. Since \async{} measurements are correlated, we set $\mbf{Q}$ with $1$s along the diagonal and $1/3$ along the off-diagonals. The unknown location of interest is
$$\best~=~\{\loc_{0} \} = \left\{ [5,~5]^\T \right\}.$$
We use a set $\seq$ that consists of four sequences: $\sseq_0~=~\{6,4,5,3\}$, $\sseq_1~=~\{3,6,4,5\}$, $\sseq_2~=~\{5,3,6,4\}$ and $\sseq_3~=~\{4,5,3,6\}$, following the scheme described in \cite{dwivedi2012scheduled}.

Similar to \tdoa, the Huber method is not readily applicable to \async{}. We therefore compare only the standard nonlinear least-squares method \eqref{eq:erm} and the proposed method \eqref{eq:main optimization problem}.

\begin{figure*}[t!]
    \centering
    \begin{subfigure}{0.45\linewidth}
    \includegraphics[width=0.9\linewidth]{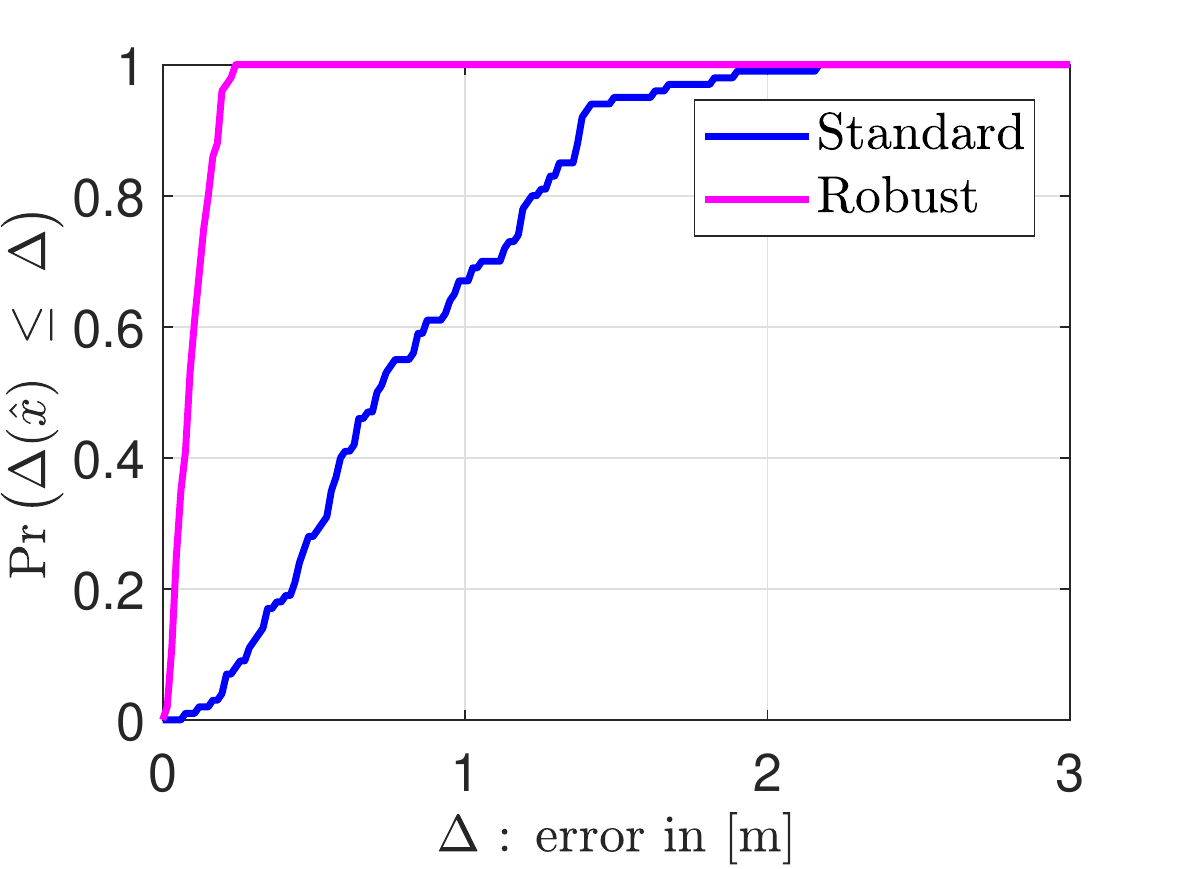}
    \caption{\cdf{}}
    \label{fig:ASYNC_CDF.pdf}
    \end{subfigure}
    \begin{subfigure}{0.45\linewidth}
    \includegraphics[width=0.9\linewidth]{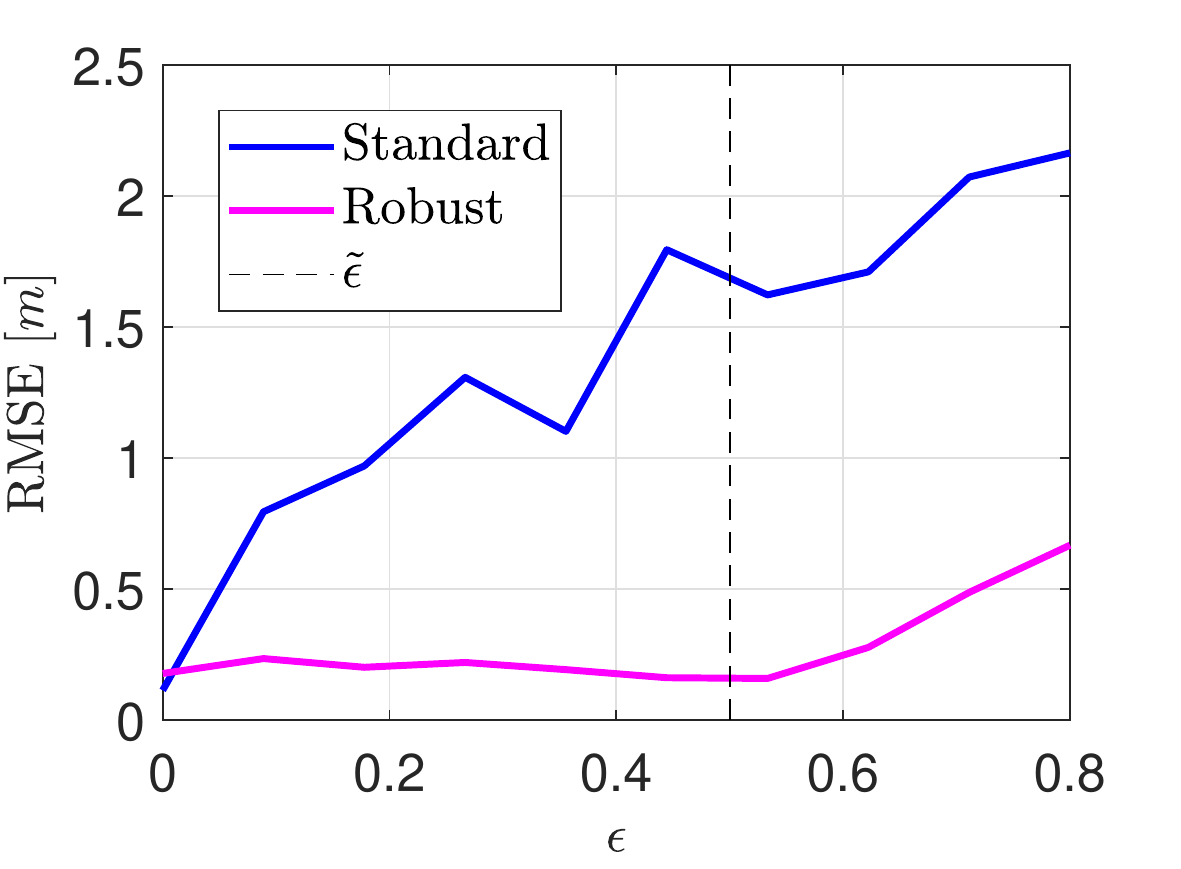}
    \caption{\rmse{}}
    \label{fig:ASYNC_RMSE_eps}
    \end{subfigure}
    \caption{Performance and sensitivity in \async{}. (a) Cumulative distribution functions of localization errors $\Delta(\widehat{\loc})$ of target node in Figure \ref{fig:3gpp all nodes} using $100$ Monte Carlo runs. Unknown corruption fraction was $\corr~=~15\%$ and the upper bound used in the robust method was set to $\corrub~=~20\%$. (b) Root-mean square error in [m] as a function of $\corr$ for the target node using standard and robust methods. Results based on $50$ Monte Carlo simulations. For the proposed robust method the upper bound was $\corrub~=~50\%$.}
    \label{fig:ASYNC}
\end{figure*}

Figure \ref{fig:ASYNC_CDF.pdf} shows the \cdf{}s of the localization error $\Delta(\widehat{\loc})$. The characteristics are similar to those in Figure~\ref{fig:toa ccdf}. The sensitivity to the unknown corruption fraction $\corr$ is  shown in Fig.~\ref{fig:ASYNC_RMSE_eps} where \rmse{} is plotted versus $\corr$ for a very conservative upper bound $\corrub~=~50\%$ in Algorithm \ref{algo:robust local}. It can be seen that the proposed method also robustifies self-localization in \async. 

\begin{figure*}[t!]
    \centering
    \begin{subfigure}{0.3\linewidth}
    \includegraphics[width=0.9\linewidth]{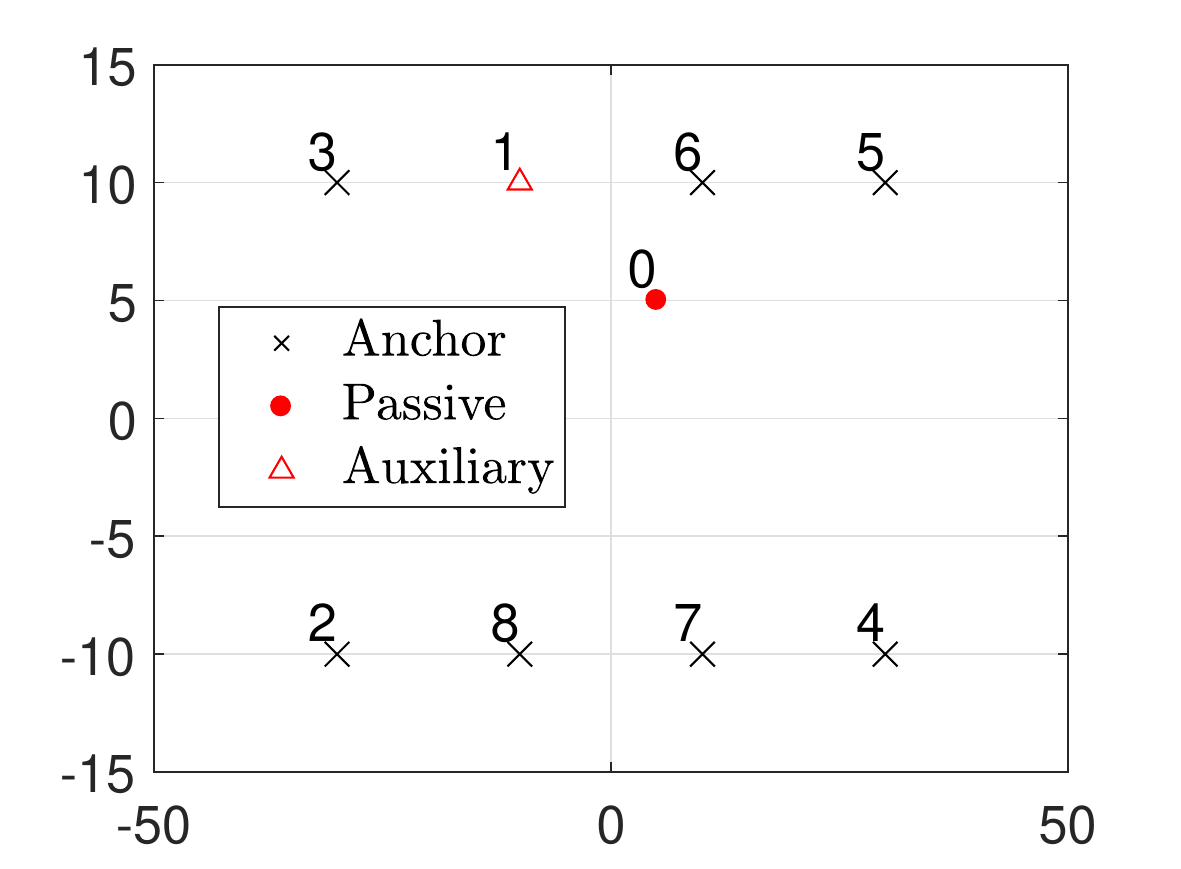}
    \caption{}
    \label{fig: passive and aux setup}
    \end{subfigure}
    \begin{subfigure}{0.3\linewidth}
    \includegraphics[width=0.9\linewidth]{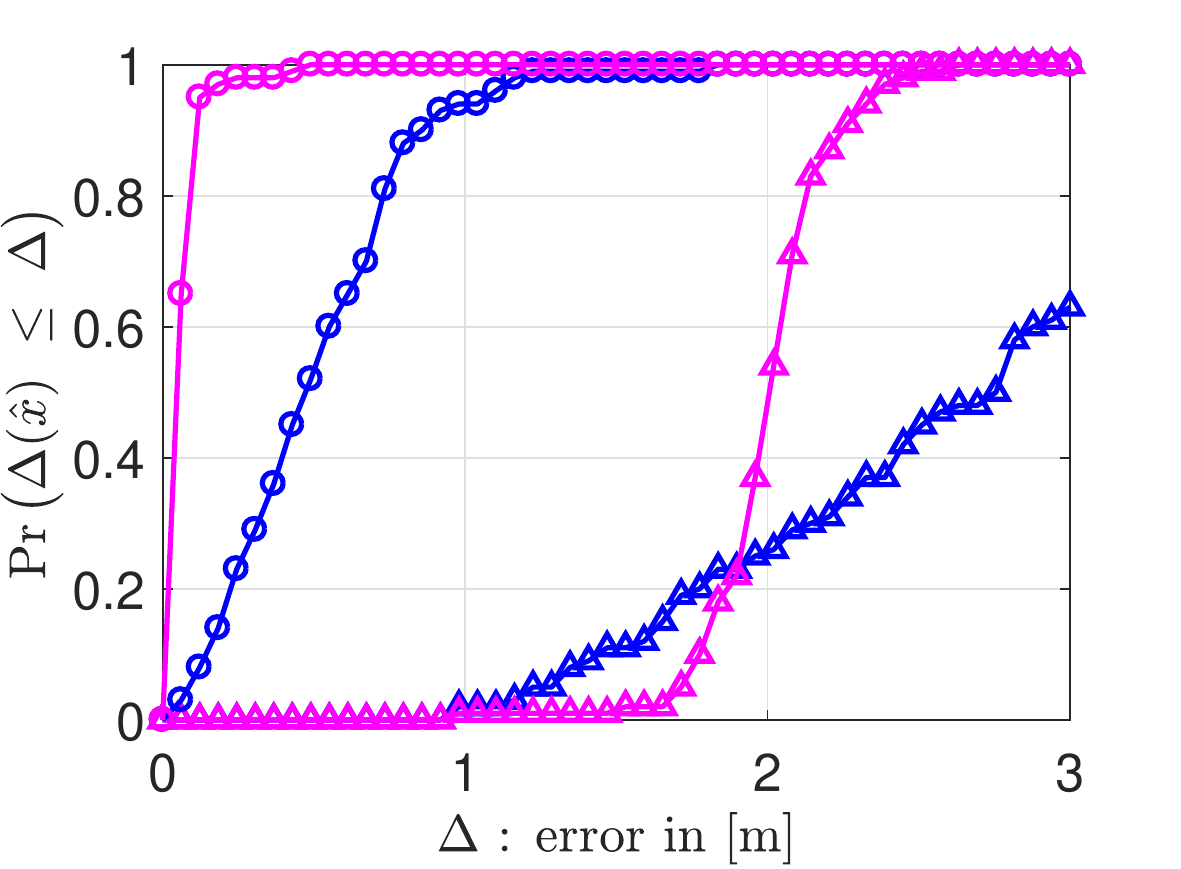}
    \caption{}
    \label{fig: box plot}
    \end{subfigure}
    \begin{subfigure}{0.3\linewidth}
    \includegraphics[width=0.9\linewidth]{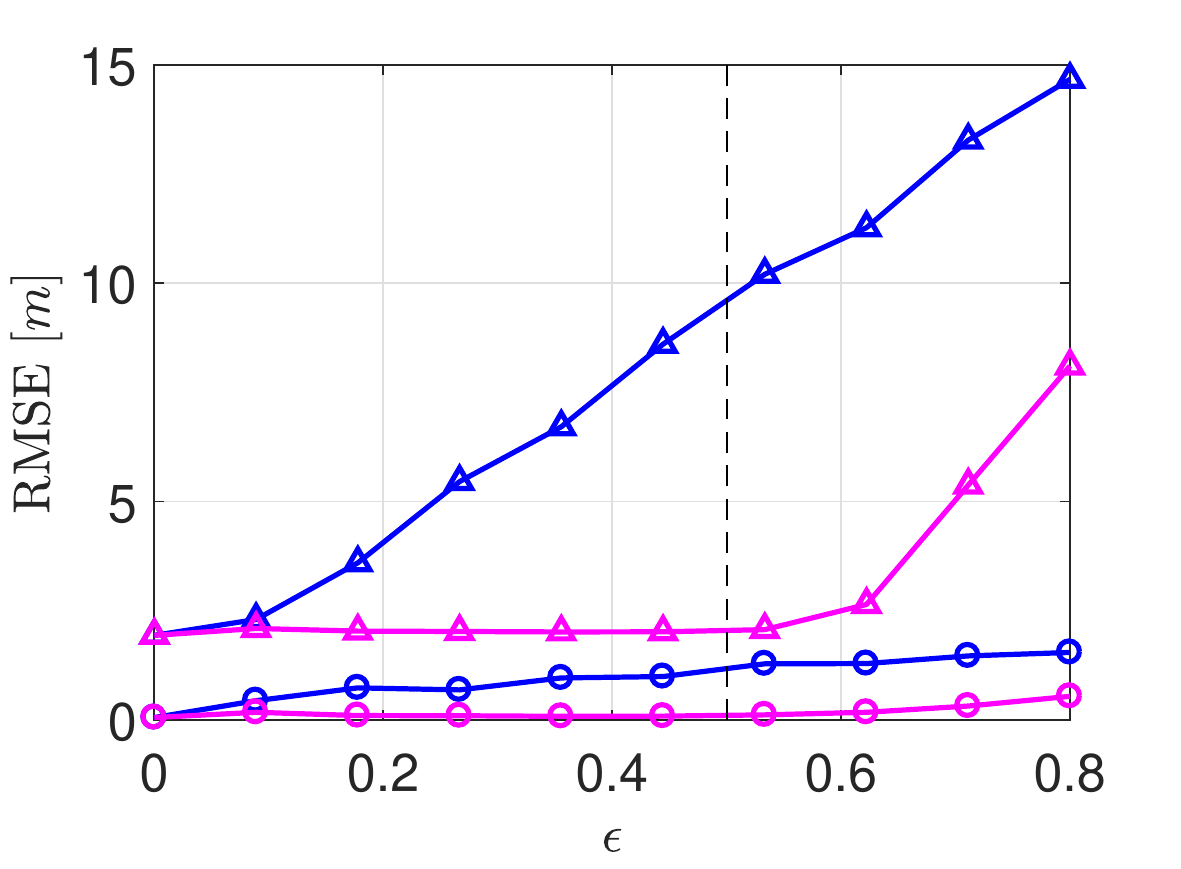}
    \caption{}
    \label{fig: mse vs eps}
    \end{subfigure}
    \caption{Performance and sensitivity in \async{} when localizing additional auxiliary node. (a) Configuration of wireless network. (b) \cdf{} of localization errors $~\Delta(\widehat{\loc})$ using $100$ Monte Carlo runs for auxiliary (triangle) and passive node (circle), using the standard (blue) and robust (magenta) methods, respectively. Unknown corruption fraction is $\corr~=~15\%$ and $\corrub = 20\%$. (c) Root mean square error as a function of the fraction of corrupted data $\corr$ using standard (blue) and robust (magenta) method, for passive (circles) and auxiliary (triangles) nodes, respectively. Results based on $50$ Monte Carlo simulations. $\corrub~=~50\%$ indicated by the black-dashed line. }
    \label{fig:passive and aux}
\end{figure*}

Finally, we consider a more challenging wireless network configuration, where one anchor node is replaced by an

auxiliary node ($\Na~=~1$) at an unknown location as shown in Figure~\ref{fig: passive and aux setup}. The goal is to  passively localize auxiliary nodes using asynchronous anchor nodes \cite{dwivedi2012scheduled,zachariah2014schedule,cavarec2017schedule} in adverse \nlos{} conditions. The locations of interest are 
$$\best~=~\{\loc_0,~\loc_1\}~=~\left\{[5,~5]^\T, [-10,~10]^\T \right\}.$$

We use a set $\seq~=~\{\sseq_0,~\sseq_1\}$, where $\sseq_{0} = \{2, 3, 4, 5, 6, 7, 8, 2, 3, 4, 5, 6, 7, 8\}$ and \\$\sseq_{1} = \{2, 1 , 3, 1, 4, 1, 5, 1, 6, 1, 7, 1, 8, 1\}$. The first sequence set $\sseq_0$ involves only the anchor nodes and ensures that the passive node location is identifiable. The second sequence $\sseq_1$ ensures that the auxiliary node location is identifiable, see \cite{zachariah2014schedule}.

Figure \ref{fig: box plot} summarizes the distribution of localization errors  $\Delta(\widehat{\loc})$ for the passive and auxiliary nodes. Note that this is a harder problem that involves two unknown locations $\loc_0$ and $\loc_1$. Moreover, when estimating $\loc_1$ the difficulty is compounded. In Figure~\ref{fig: mse vs eps} we see that the localization 

performance degrades  rapidly under \nlos{} conditions using the standard method, whereas the proposed method is resilient.

\section{Conclusion}
We have developed a robust method for localization of nodes in  wireless networks using timing-based measurements. The method is applicable to a wide range of localization technologies. Its  robustness properties against data corrupted due to non-ideal and non-line-of-sight signal conditions are demonstrated using three different configurations: \toa{}, \tdoa{} and \async. The proposed method does not make any assumptions on the form of corrupting noise distribution, and  only requires an upper bound on the fraction of corrupted data.



\bibliographystyle{unsrtnat}

\end{document}